%% file: main.tex
\documentclass[sigconf, screen, natbib=false, nonacm]{acmart}
\AtBeginDocument{%
  }

\setcopyright{acmlicensed}

\copyrightyear{2026}
\acmYear{2026}
\acmDOI{XXXXXXX.XXXXXXX}
\acmConference[SAC'26]{The 41st ACM/SIGAPP Symposium on Applied Computing}{March 23--27, 2026}{Thessaloniki, Greece}
\acmISBN{979-X-XXXX-XXXX-X/26/03}

\input{sections/preamble}

\RequirePackage[
  datamodel=acmdatamodel,
  style=acmnumeric,
  ]{biblatex}
\usepackage{amsthm}

\newtheorem*{definition}{Definition}

\begin{document}

\title{From See to Shield: ML-Assisted Fine-Grained Access Control for Visual Data
}
\renewcommand{\shorttitle}{ML-Assisted Access Control}

\author{Mete Harun Akcay}
\authornote{Work done during internship at Nokia Bell Labs}
\affiliation{%
  \institution{Abo Akademi University, Nokia Bell Labs, Finland}
  \country{}
}
\email{meteharun.ackay@abo.fi}

\author{Buse Gul Atli}
\affiliation{%
 \institution{Linköping University, Nokia Bell Labs, Sweden}
 \country{}
 }
\email{busega@acm.org}

\author{Siddharth Prakash Rao}
\affiliation{%
  \institution{Nokia Bell Labs, Finland}
  \country{}
  }
\email{sid.rao@nokia-bell-labs.com}

\author{Alexandros Bakas}
\affiliation{%
  \institution{Nokia Bell Labs, Finland}
  \country{}
}
\email{alexandros.bakas@nokia-bell-labs.com}

\renewcommand{\shortauthors}{Akcay et al.}

\begin{abstract}
As the volume of stored data continues to grow, identifying and protecting sensitive information within large repositories becomes increasingly challenging, especially when shared with multiple users with different roles and permissions. This work presents a system architecture for trusted data sharing with policy-driven access control, enabling selective protection of sensitive regions while maintaining scalability. The proposed architecture integrates four core modules that combine automated detection of sensitive regions, post-correction, key management, and access control. Sensitive regions are secured using a hybrid scheme that employs symmetric encryption for efficiency and Attribute-Based Encryption for policy enforcement. The system supports efficient key distribution and isolates key storage to strengthen overall security. To demonstrate its applicability, we evaluate the system on visual datasets, where Privacy-Sensitive Objects in images are automatically detected, reassessed, and selectively encrypted prior to sharing in a data repository. Experimental results show that our system provides effective PSO detection, increases macro-averaged F1 score (5\%) and mean Average Precision (10\%), and maintains an average policy-enforced decryption time of less than 1 second per image. These results demonstrate the effectiveness, efficiency and scalability of our proposed solution for fine-grained access control.

\end{abstract}





\settopmatter{printfolios=true}
\maketitle

\input{sections/intro}
\input{sections/related_work}

\input{sections/method}

\input{sections/experimental}
\input{sections/results}
\input{sections/security}

\input{sections/conclusion}
\printbibliography


\end{document}

%% file: sections/preamble.tex
\usepackage{graphicx}
\usepackage{subcaption}

\usepackage{makecell}
\usepackage{fontawesome}
\usepackage{multirow}
\usepackage{arydshln}

\newcommand{\PSO}{PSO}

\usepackage{todonotes}

\newcommand{\boldtitle}[1]{\textbf{#1}}

%% file: sections/intro.tex
\section{Introduction}

The rapid growth of data-driven digital systems and applications has led to the collection and storage of large volumes of sensitive data, ranging from personally identifiable information (PII) to financial transactions and healthcare records. Managing and regulating access to this information has become a critical challenge for individuals and organizations. Beyond protection against external threats, effective access control is a cornerstone of data security, ensuring that only authorized users, applications, and processes can interact with sensitive information. Access control mechanisms regulate and restrict access to resources, systems, or physical areas, typically granting permissions to users based on predefined roles and policies. 
%
%
For example, access to medical records is limited to healthcare professionals, surveillance footage is restricted to security personnel, and digital copies of sealed judicial records are provided only to lawyers and judges. 
While dealing with such semantically nuanced restrictions, access control mechanisms must operate both reliably and efficiently. The consequences of inadequate access control mechanisms can result in serious financial, reputational, and operational damage as seen in major data breach incidents\footnote{https://www.ibm.com/reports/data-breach}. 
Complementing this reality, regulatory frameworks (e.g., the GDPR~\cite{GDPR2016a} and the EU AI Act~\cite{AIAct24}) have begun to establish binding obligations for organizations to secure sensitive information. 
The demonstrated harm of breaches in practice, combined with the tightening demands of regulation, points towards an urgent need for robust access control mechanisms that can adapt to the scale and complexity of modern digital ecosystems.

Access control mechanisms are usually used in conjunction with data classification and encryption, each with its own limitations. Data classification is often performed manually by end users, who may not fully understand the risks of misclassification, resulting in amplified human errors and inconsistencies when scaled. Machine Learning (ML) has great potential for data classification. 
However, ML methods may have inherent limitations, such as limited performance due to domain adaptation and the scarcity of properly annotated datasets that contain public and varying levels of private information. 
Encryption, on the other hand, is often implemented independently and in isolation from access control. Also, encryption is often applied coarsely by treating files or databases as a single entity. This ``all-or-nothing'' operational model limits usability and creates bottlenecks when selective portions of the data require protection. Thus, the limitations of data classification and encryption, when combined with those of access control, tend to reinforce one another's weaknesses rather than complementing their strengths.

In this paper, we present a system architecture that enables fine-grained access control to protect sensitive information within files in a selective manner. We demonstrate the feasibility of our solution by applying it to visual datasets. Our choice is driven by the fact that such datasets are inherently heterogeneous in nature and often require classification beyond the file level, extending to granularity (such as specific regions, objects, or attributes) of the visual content of individual files~\cite{zhao2025visual}. Our proposed system utilizes ML techniques to determine where to apply access protection selectively by detecting Privacy Sensitive Objects (PSOs) within the file and classifying these objects based on their varying levels of sensitivity. Regarding how to apply access protection, we make a critical observation that it is commonplace to use blurring- or pixelation-based redaction to protect sensitive information in images. These methods are less secure because redaction only transforms the underlying content into a distorted, lossy representation that is vulnerable to reconstruction attacks, in which an adversary can accurately reconstruct the redacted parts~\cite{bland2022story}. To address this, our solution employs cryptographic protection that not only provides stronger protection against reconstruction attacks, but also enables controlled reconstruction by design, allowing authorized users to recover protected regions securely through decryption.

\boldtitle{Contributions}: We present a system that orchestrates ML and cryptographic techniques for fine-grained access control, and contribute in the following directions.

\textbf{PSO Detection}:  We develop an ML-based component for PSO detection and scoring. The framework is built on results from \textit{performance benchmarking of 13 off-the-shelf ML models} for detecting sensitive information across different modalities (i.e., textual, visual, or multimodal) in visual images. Also, we introduce \textit{two novel post-correction methods}, namely Context-Aware Post Correction (CAPC) and Post-BERT, that leverage semantic and textual cues to refine predictions. Using benchmarking to identify the best-performing off-the-shelf ML model for each modality and post-correction to enhance contextual accuracy, our work can be viewed as a step toward building a high-level ensemble baseline that integrates modality-specific models for comprehensive PSO detection.

\textbf{PSO Protection}: We develop a cryptographic protection component that complements PSO detection through a \textit{hybrid encryption scheme} and an\textit{ optimized policy design}. The hybrid scheme combines symmetric encryption to encrypt the PSOs and Attribute-Based encryption (ABE) to encrypt the symmetric keys. While symmetric encryption offers performance and scalability, ABE integrates encryption with fine-grained access control within a single mechanism without external enforcement. Also, the policy-driven nature of ABE makes it a suitable candidate for seamless integration with widely deployed access control frameworks~\cite{parkinson2022survey}. 
The optimized policy design for ABE reduces operational overhead and enhances scalability by enabling each authorized user to perform a single ABE decryption to efficiently recover all symmetric keys associated with their access rights and by avoiding a naive approach of multiple decryptions for different portions of the data.

\boldtitle{Overview of results}: 
We evaluate the performance of our proposed system on images containing multiple PSOs with varying sensitivities. 
The PSO detection component demonstrates an effective performance, with our post-correction methods improving the macro-averaged F1 score of the best text classification model by 5\% and the mean Average Precision (mAP) of the best object detection model by 10\%, while maintaining an acceptable Mean Intersection over Union (mIoU) score of up to 80.7\% across segmentation models. Each image in our dataset contains an average of 5.6 PSOs, and the complete detection pipeline, including scoring and metadata generation, runs at an average of 6.01 seconds per image. On the other hand, the PSO protection component demonstrates efficiency by using an optimal, bounded number of keys independent of the number of PSOs or users. The computational overhead is efficient, with~$\sim$11 seconds for encryption and~<1 second for decryption even in worst-case scenarios. The storage overhead grows linearly with dataset size and is inherently limited by the maximum number of PSOs detectable per image, ensuring predictable and sustainable scalability. These results collectively demonstrate the feasibility of enforcing fine-grained, selective access control without compromising efficiency or scalability.

%% file: sections/related_work.tex
\section{Related Work}
\label{sec:related}

\paragraph{Object Detection, Segmentation \& Character Recognition:} Computer vision applications heavily employ machine learning methods for three core tasks: object detection, segmentation, and optical character recognition (OCR). Object detection~\cite{yolov8, ren2015faster} is used to locate, identify, and classify objects within an image, while segmentation~\cite{he2017mask,sam,xie2021segformer} labels each pixel with a pre-defined class. Semantic segmentation labels every pixel in an input image without differentiating between individual objects. In contrast, instance segmentation provides pixel-level labels while also identifying each object separately. State-of-the-art techniques for object detection and segmentation have achieved impressive results in identifying visual components. 
However, these methods are not equipped to extract text. In contrast, OCR systems~\cite{paddleocr, smith2007tesseract} have been developed to recognize text within images. OCRs are widely used to understand scanned documents or identity verification. The combination of these methods are used in various applications, including pose detection, video captioning, and scene graph prediction, and as a fundamental component in self-driving cars. 


\paragraph{Private Content \& Privacy Risk Scoring in Visual Data:} Orekondy et al.~\cite{Orekondy2018Connecting} introduced the first approach for automated redaction of private content from images. They derived a dataset from~\cite{orekondy17iccv} by selecting images with privacy-sensitive regions that can be localized for redaction, providing annotations for private objects. They also evaluated semantic segmentation methods for automated redaction via masking. Similarly, Gurari et al.~\cite{Gurari2019VizWiz} released a visual privacy recognition dataset, which contains images captured by people with visual impairments, and evaluated unintentional privacy leaks in visual question answering tasks. Building on the notion of visual privacy risk score in~\cite{orekondy17iccv}, Chen et al.~\cite{Chen2021Privattnet} used LSTMs with attention maps to estimate the privacy risk of an entire image. However, their approach does not explicitly localize sensitive objects. 
More recently, Tay et al.~\cite{Tay2024PrivObfNet} combined attention maps with weakly supervised semantic segmentation to predict privacy scores, identify categories of sensitive objects, and generate masks for obfuscation. Although the method in~\cite{Tay2024PrivObfNet} improves attention quality compared to previous segmentation models, it focuses primarily on a single object and roughly covers all textual areas while failing to localize text objects. Tseng et al.~\cite{Tseng2025BIV} extended the work in~\cite{Gurari2019VizWiz} by releasing a new dataset, which contains images with segmented private objects, tailored for the localization of sensitive regions. They also benchmarked several few-shot object detection and segmentation models on this dataset. However, their dataset typically contains only a single annotated privacy sensitive object, reducing the need for fine-grained, multilevel obfuscation mechanisms.

\paragraph{Secure Data Sharing and Access Control:} Prior work has explored privacy-preserving data sharing in cloud and IoT environments, including revocable and collaborative ABE schemes~\cite{4223236,bakas2019modern,Xu2018Finegrained,Xue2019Attribute} for dynamic user groups, SeDaSC~\cite{Ali2017SeDaSC} which counters insider threats using key shares managed by a trusted third party, and IoT-focused approaches~\cite{cryptoeprint:2020/176,Mollah2017Secure} such as Symmetric Searchable Encryption for encrypted search and offloading of heavy security operations to edge servers. Other specialized solutions include SecRCNN~\cite{Liu2022Privacy}, a lightweight privacy-preserving Faster R-CNN framework for medical images using additive secret sharing and edge computing, and a user-centric data space~\cite{Robert2025Unlinkable} combining differential privacy with fine-grained access control for unlinkable data sharing via a central intermediary. In contrast, our work embeds fine-grained access control directly into the cryptographic layer via a hybrid Attribute-Based Encryption scheme, efficiently protecting privacy-sensitive objects in visual data without relying on trusted external intermediaries, interactive protocols, or heavy computation on end devices.

\section{Preliminaries}
\label{sec:background}



\paragraph{Sensitivity Scores \& Groups:}To support fine-grained control, each privacy-sensitive object (PSO) is assigned a sensitivity score on a continuous scale $[\alpha, \beta]$, where $\alpha$ denotes the lowest and $\beta$ the highest degree of sensitivity. These scores quantify the relative privacy risk associated with individual PSOs and serve as the basis for grouping and policy assignment. After scoring, PSOs are assigned discrete sensitivity groups to simplify policy enforcement. The number of groups $n$ is configurable by the system administrator, organizational policies, privacy regulations, or user-defined requirements. 
The interval $[\alpha,\beta]$ is partitioned into bins $n$, each corresponding to a distinct sensitivity group $G_1, \dots, G_n$. The boundaries of each group $[\alpha_\ell, \beta_\ell)$, define the sensitivity thresholds. Groups with higher score ranges represent more sensitive information (similar to document classification levels), and therefore require stricter access control and stronger cryptographic protection.

\paragraph{Notation:}
Policies are denoted by $P$ and are constructed as logical combinations of attributes. The attributes of a user $u$ are represented as a vector $\mathbf{a} = [a_1, \dots, a_n]$, where each $a_i$ corresponds to an attribute such as a user’s \textit{role} (e.g., \textit{doctor}, \textit{nurse}, \textit{manager}). A policy $P$ is said to be satisfied by a user’s attribute set $\mathbf{a}$ if $P(\mathbf{a}) = \text{True}$. The output $y$ of a probabilistic algorithm $\mathsf{A}_P$ on input $x$ is denoted by $y \leftarrow \mathsf{A}_P(x)$, while the output $y$ of a deterministic algorithm $\mathsf{A}_D$ on input $x$ is denoted by $y := \mathsf{A}_D(x)$.

Our work relies on a hybrid encryption scheme between two different cryptographic primitives. More precisely, a symmetric key encryption scheme $\mathsf{SKE}$ and an attribute-based encryption scheme $\mathsf{ABE}$.
$\mathsf{ABE}$ extends public-key encryption by associating ciphertexts with access policies and users with attribute-based secret keys. A user can decrypt a ciphertext only if their attributes satisfy the policy, providing fine-grained access control. However, ABE can be computationally intensive for large datasets, which has motivated the use of hybrid approaches where ABE is applied primarily to encrypt symmetric keys, while $\mathsf{SKE}$ handles bulk data efficiently. More formally:

\begin{definition}[Symmetric Key Encryption]
    A symmetric-key encryption scheme $\mathsf{SKE}$ for a message space $\mathcal{M}$ and a target space $\mathcal{C}$ consists of three polynomial-time algorithms $(\mathsf{Gen, Enc, Dec})$ such that:
    \begin{itemize}
        \item $\mathsf{SKE.Gen}(1^\lambda)$: Takes as input a security parameter $\lambda$ and outputs a symmetric key $\mathsf{K}$.
        \item $\mathsf{SKE.Enc}(\mathsf{K}, m)$: Takes as input a symmetric key $\mathsf{K}$ and a plaintext $m \in \mathcal{M}$ and outputs an encrypted message $c\in \mathcal{C}$ 
        \item $\mathsf{SKE.Dec}(\mathsf{K}, c)$: Takes an input a symmetric key $\mathsf{K}$ and ciphertext $c\in \mathcal{C}$, and outputs a plaintext $m\in\mathcal{M}$
    \end{itemize}
\end{definition}

\begin{definition}[Attribute-Based Encryption]
    An attribute-based encryption scheme $\mathsf{ABE}$ for a message space $\mathcal{M}$ and a target space $\mathcal{C}$ consists of four polynomial-time algorithms ($\mathsf{Gen, Enc, KeyGen, Dec}$) such that:
    \begin{itemize}
        \item $\mathsf{ABE.Gen}(1^\lambda)$: Takes as input a security parameter $\lambda$ and outputs a master public/private key pair $\mathsf{(mpk, msk)}$.
        \item $\mathsf{ABE.Enc}(\mathsf{mpk, P, m})$: Takes as input a master public key $\mathsf{mpk}$, an access policy $P$ and a plaintext $m \in \mathcal{M}$ and outputs a ciphertext $c_P \in \mathcal{C}$, bound to the policy $P$.
        \item $\mathsf{ABE.KeyGen}(\mathsf{msk}, \mathbf{a})$: Takes as input a master secret key $\mathsf{msk}$ and a list of attributes $\mathbf{a}$, and outputs a decryption key $\mathsf{sk}$.
        \item $\mathsf{ABE.Dec}(\mathsf{sk}, c_P)$: Takes as input a secret decryption key $\mathsf{sk}$ and a ciphertext $c_P \in \mathcal{C}$, and outputs $m \in \mathcal{M}$ iff $P(\mathbf{a}) = True$.
    \end{itemize}
\end{definition}

%% file: sections/method.tex
\section{Proposed Architecture}
\label{sec:Architecture}

\begin{figure*}[t]
\centering
        \includegraphics[width=1.0\textwidth]{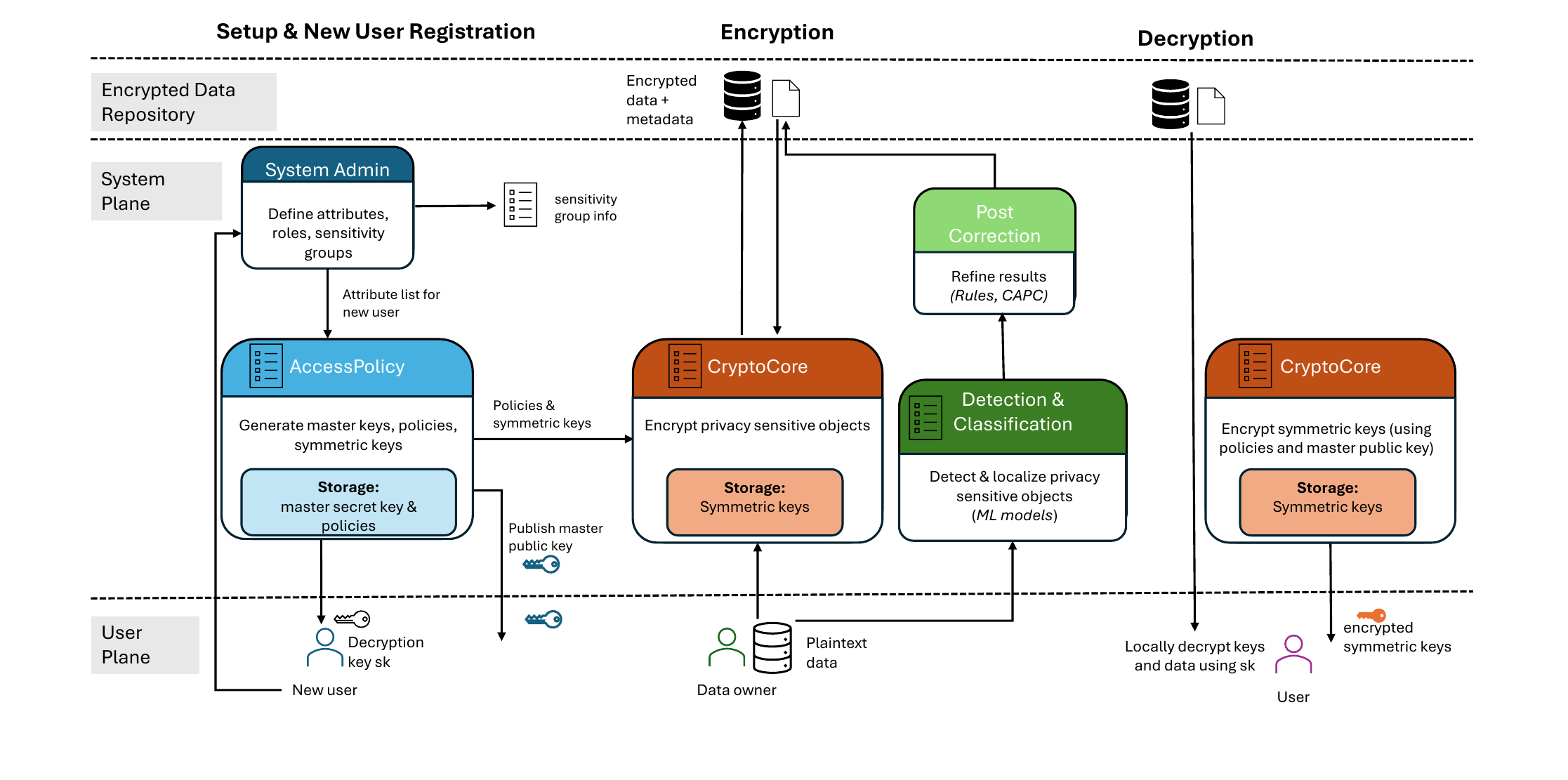}
    \caption{High-level system architecture illustrating three operational phases: (a) registration, (b) data encryption, and (c) data decryption. The system includes key actors (system administrator, data owners, and users) and four core modules (AccessPolicy, Detection, Post-Correction, and CryptoCore). Sensitive data provided by the owner is processed, labeled, and encrypted before storage. Users receive personalized decryption keys and can only access protected content if authorized by the embedded policy.}
    \label{fig:architecture}
\end{figure*}

Our proposed architecture (Figure~\ref{fig:architecture}) is organized into three logical layers: \textit{System Plane}, \textit{User Plane}, and \textit{Encrypted Data Repository}. This separation establishes clear functional boundaries between system components, user interactions, and data storage locations.

The \textit{System Plane} comprises four core modules, each responsible for specific functionalities. These modules interact within the system plane, and also coordinate with other logical layers. At a high level, the \textit{AccessPolicy} module (refer to Section~\ref{subsec:access_policy_module}) is responsible for generating cryptographic keys and provisioning access control policies based on the inputs from a \textit{System Admin} about user attributes or roles and sensitivity group descriptions. The \textit{Detection \& Classification} and \textit{Post-correction} modules (refer to Sections~\ref{subsec:detection_module} and ~\ref{subsec:correction_module}) process raw (i.e., unencrypted) input image data to identify and categorize PSOs. The \textit{CryproCore} module (refer to Section~\ref{subsec:crypto_module}) is responsible for handling encryption and decryption operations. While the CryptoCore module handles the storage and management of symmetric keys, the same for ABE decryption keys are handled by the AccessPolicy module. 

The \textit{User Plane} represents how \textit{end-users} and \textit{data owners} interact with the system during three operational phases: (i) \textit{Setup \& New-user registration}, where the system admin defines attributes, roles, sensitivity groups, as well as registers new users while the AccessPolicy module generates all cyrptographic keys (ii) \textit{Encryption}, where data owners supply plaintext data that traverse the Detection \& Classification, Post-Correction, and CryptoCore modules before being securely stored, and  (iii) Decryption, where authorized users retrieve encrypted data and decrypt only the portions permitted by their assigned access policies.


The \textit{Encrypted Data Repository} stores the encrypted data and associated metadata provided by the system plane. It is location-agnostic and can reside on the enterprise server or cloud platform, depending on the deployment requirements. Users can access encrypted files, but they require the correct keys to decrypt them.

\subsection{AccessPolicy Module}
\label{subsec:access_policy_module}

The AccessPolicy module forms the backbone of our system's privacy and access control framework. It receives the sensitivity group definitions (i.e., the thresholds for each group), the roles defined by the system administrator, and a list of authorized users with their associated attributes. This module is responsible for:


\begin{enumerate}
    \item Defining the policies for each sensitivity group.
    \item Generating all cryptographic keys required by the system.
    \item Providing the CryptoCore Module with the defined policies and the symmetric keys required to perform the encryption of sensitive data.

\end{enumerate}


\paragraph{Key Generation:}  
For each user-specified sensitivity group $G_\ell$, where $\ell \in \{1, \dots, L\}$, the AccessPolicy Module generates a symmetric key $\mathsf{K}_\ell$ by executing $\mathsf{K}_\ell \leftarrow \mathsf{SKE.Gen}(1^\lambda)$. The symmetric key for sensitivity group $G_1$ is simply $\mathsf{K}_{G_1}$, while for each group $G_i$ with $i > 1$, the key is defined as: 
\[\mathsf{K}_{G_i}=\mathsf{K}_i \| \mathsf{K}_{i-1} \| \dots \| \mathsf{K}_1. 
\tag{1}\label{eq:1} 
\]
This structure ensures that a user that has access to group $G_i$ can also access data in all groups $G_j$ where $j < i$. 

AccessPolicy also generates a master public/private key pair for the $\mathsf{ABE}$ scheme by running $(\mathsf{mpk}, \mathsf{msk}) \leftarrow \mathsf{ABE.Gen}(1^\lambda)$.It publishes $\mathsf{mpk}$, while $\mathsf{msk}$ remains private and never leaves the module. Finally, we assume that the AccessPolicy Module has access to a list of authorized users in the system, along with their associated attributes.  For each registered user $u$, an $\mathsf{ABE}$ decryption key $\mathsf{sk}_u$ is derived from $\mathsf{msk}$ according to the user’s attribute set $\mathbf{a}_u$ via $\mathsf{sk}_u \leftarrow \mathsf{ABE.KeyGen}(\mathsf{msk}, \mathbf{a}_u)$.

\paragraph{Policy Definition \& Encryption Delegation:}  
Once the symmetric keys are generated, the AccessPolicy Module defines a distinct policy for each sensitivity group. These policies, together with the ABE master public key $\mathsf{mpk}$, are then passed to the CryptoCore Module, which performs the encryption of the symmetric keys $\mathsf{K}_1, \dots, \mathsf{K}_L$, producing ciphertexts $c_{P_1,\mathsf{K}_1}, \dots, c_{P_L,\mathsf{K}_L}$.

For each sensitivity group $G_\ell$, a monotone policy is constructed as a disjunction of attributes:
\[
P_\ell = a_1 \lor a_2 \lor \dots \lor a_r,
\tag{2}\label{eq:2} 
\]
where each $a_j$ represents an attribute that a user may have. Some of these attributes may satisfy the policy of the group $G_\ell$ but \textbf{not} policies of groups $G_k$ with $k > \ell$, while other attributes may satisfy the policies of multiple groups simultaneously. This design ensures that any user that holds at least one attribute that satisfies the policy $P_\ell$ can decrypt the symmetric key for the group $G_\ell$. Furthermore, users who have attributes that satisfy the policies of the higher-sensitivity groups can also access all the lower-sensitivity groups. Table~\ref{tab:ABEDec} presents a toy example that demonstrates which users have the ability to decrypt particular sensitivity groups.

\begin{table*}[h!]
\centering
\caption{Example of ABE decryption capabilities for $L=3$ sensitivity groups and $\ell \in \{1, 2, 3\}$.}
\label{tab:ABEDec}

\begin{tabular}{|c|c|c|}
\hline
\textbf{Sensitivity Group} & \textbf{Policy $P_\ell$} & \textbf{Attributes Allowed to Decrypt} \\
\hline
1 & $a_1 \lor a_2 \lor a_3$ & Users with $a_1$, or $a_2$, or $a_3$ \\
\hline
2 & $a_2 \lor a_3$ & Users with $a_2$ or $a_3$ \\
\hline
3 & $a_3$ & Users with $a_3$ \\
\hline
\end{tabular}
\end{table*}

\subsection{Detection and Classification Module}
\label{subsec:detection_module}

Given input data, this module detects, classifies and localizes privacy-sensitive objects ({\PSO}s). It extracts either the mask information or bounding-box coordinates (depending on the modality) and forwards them to the post-correction module, along with the predicted label, the confidence score of the prediction, the corresponding sensitivity score, and the mapped sensitivity group. 

Detecting {\PSO}s requires different strategies depending on the modality of the object. Using a single model is often insufficient, since {\PSO}s vary greatly in shape, appearance, and reliance on visual vs. semantic cues. Following the definition in prior work~\cite{Orekondy2018Connecting}, we categorize {\PSO}s into three groups based on the modality: 
\begin{itemize}
    \item \textit{Visual {\PSO}s} (e.g., face, handwriting, signature) that exhibit irregular shapes, colors, and textures
    \item \textit{Textual {\PSO}s} (e.g. names, dates, phone numbers) that are inherently semantic, where pixel-level information is insufficient, and detection requires character recognition
    \item \textit{Multimodal {\PSO}s} (e.g., credit cards, ID documents, tickets) that combine structural features with embedded text, requiring visual and semantic analysis
\end{itemize}

For visual {\PSO}s, segmentation models are used to extract pixel-level information, minimizing the privacy/utility trade-off. 
For textual {\PSO}s, OCR is applied to parse the image text, which is then classified using natural language processing models. Unlike visual {\PSO}s, the detection module outputs bounding-box coordinates for textual regions. For multimodal {\PSO}s, object detection models are used instead of segmentation to extract bounding-box coordinates, since segmentation may fail to label pixels belonging to critical areas (e.g., barcodes in ID cards) in the object. Finally, to refine both the localization and the predicted label of textual and multimodal {\PSO}s, the output is passed to the Post-Correction module.

\subsection{Post-Correction Module}
\label{subsec:correction_module}

Post-Correction complements the detection module by enabling finer-grained and context-aware classification for both textual and multimodal {\PSO}s.

For textual {\PSO}s, the semantics of an isolated word is often insufficient. For example, distinguishing between a generic date and a birthdate requires contextual information from nearby text. Therefore, this module corrects the classification and localization results of textual {\PSO}s with rule-based adjustments, informed by spatially adjacent cues in the image, as follows. 
%
%
\begin{itemize}
    \item If a text contains a \textit{temporal identity} cue (e.g., ``dob'', ``born'', or ``birthday''), and if the predicted label of the closest object is \emph{date}, that prediction is updated as \textit{birthdate}.
    \item If a text contains a \textit{personal identity} cue (e.g., ``name'', ``surname'', or ``alias''), and if the predicted label of the closest object is \textit{safe}, that prediction is updated as \textit{name}. 
    \item  If a text contains a \textit{location} cue (e.g., ``office'' or ``city''), and if the predicted label of the closest object is \textit{safe}, that prediction is updated as \textit{place}.
\end{itemize}
%
%
While rules above adequately capture the semantics or contextual relations of the images in our use case, real-world deployments may have to accommodate broader or generic rule-based refinements.

For multimodal {\PSO}s, this module improves object detection results with text analysis. Object detection in previous module can not distinguish visually similar multimodal PSOs such as driver's license vs. student ID. Therefore, we design the \emph{Context-Aware Post-Correction (CAPC)} algorithm, which applies OCR to the detected region, classifies the extracted text with a natural language processing model, and updates the final predicted label accordingly to differentiate between visually similar objects. 

After detection and post-correction, a metadata file is generated to facilitate encryption and decryption. The metadata includes reassigned label, annotation at the pixel or bounding-box level, confidence score for the predicted label, sensitivity score, and the associated sensitivity group per each detected {\PSO} in every image in the dataset designated for encrypted data repository. 
If AccessPolicy module modify the rules later (e.g., adjusting the number of sensitivity groups or modifying thresholds) or introduce new rules (e.g., encrypting {\PSO}s only when the confidence score exceeds a specified value), these updates can be applied directly to the metadata without rerunning the detection and post-correction modules. 

\subsection{CryptoCore Module}
\label{subsec:crypto_module}

The CryptoCore module provides fine-grained access control over privacy-sensitive objects by combining ABE with symmetric encryption. This hybrid design ensures that sensitive content can be flexibly shared between different users while maintaining strict confidentiality guarantees.

\paragraph{PSO Encryption:}

As a first step, the CryptoCore module retrieves the symmetric keys ${\mathsf{K}_{G_\ell}}$ associated with each sensitivity group $G_\ell$, $\ell \in \{1, \dots, L\}$. Using the metadata generated by the Post-Correction module, 
CryptoCore determines which symmetric key to use for encrypting each PSO, ensuring that the encryption process is consistent with the sensitivity groups. 
Given a single $\mathsf{PSO}_i$ belonging to a sensitivity group $\mathsf{K}_{G_{\ell}}$, CryptoCore encrypts it by computing
\[
    c_{\mathsf{PSO}_i} \leftarrow \mathsf{SKE.Enc}(\mathsf{K}_{G_{\ell}}, \mathsf{PSO}_i). 
    \tag{3}\label{eq:3} 
\]
The $\mathsf{SKE.Enc}$ algorithm utilizes only the portion of $\mathsf{K}_{G_{\ell}}$ corresponding to $\mathsf{K}_\ell$ during encryption, as stated earlier in Equation~\ref{eq:1}. We should note that all sensitive regions are processed in descending order of sensitivity, starting from the highest sensitivity groups. If
a region is already encrypted with a symmetric key corresponding to a higher sensitivity group, it is not re-encrypted with symmetric keys from lower groups, since users holding keys for higher sensitivity groups can also decrypt all content in lower groups. This approach may render PSOs belonging to lower-sensitivity groups inaccessible to certain users if the entire PSO is contained within a higher-sensitivity region. We prioritize protecting the privacy of higher-sensitivity groups, even at the cost of limiting access to lower-sensitivity objects. Additionally, this strategy eliminates redundant computations by ensuring that overlapping regions are encrypted only once.

To enforce access control, the symmetric keys $\mathsf{K}_{G_{\ell}}$ are also encrypted using ABE, thus protecting each $\mathsf{K}_{G_{\ell}}$ under a distinct access policy as specified by the AccessPolicy Module in~\ref{subsec:access_policy_module}. This mechanism binds decryption capabilities directly to user attributes, enabling flexible role- or context-based authorization. Similarly to encrypting PSOs, the module first retrieves the policies $P_\ell$ associated with each sensitivity group and proceeds by encrypting the symmetric keys using ABE. More specifically, for each symmetric key $\mathsf{K}_{G_{{\ell}}}$, the CryptoCore module computes:
\[
    c_{P{_\ell}_{\mathsf{K}_{G_{{\ell}}}}} \leftarrow \mathsf{ABE.Enc}(\mathsf{mpk}, P_\ell, \mathsf{K}_{G_\ell})
    \tag{4}\label{eq:4} 
\]

Since each symmetric key is associated with only one sensitivity group, this module performs exactly as many ABE encryptions as the number of sensitivity groups.

\paragraph{Decryption:}
Although all registered users can access the encrypted data repository, they must request the corresponding encrypted symmetric keys from the CryptoCore module. Users only need to retrieve the encrypted keys once; subsequent decryptions can be performed locally using the keys already obtained.

After receiving the encrypted symmetric keys, the ABE decryption key $\mathsf{sk}_u$  is used to attempt decryption of the per-group key ciphertexts, starting from the most sensitive one. The first successfully decrypted key determines the maximum sensitivity group that the user is authorized to access. That is, upon receiving an encrypted PSO ($c_{\mathsf{PSO}}$) and an encrypted symmetric key $c_{P{_\ell}_{\mathsf{K}_{G_{\ell}}}}$, the user first tries to decrypt the symmetric key using their ABE decryption key $\mathsf{sk}_u$. In particular, the user computes:
\[
\mathsf{ABE.Dec}(\mathsf{sk}_u, c_{P{_\ell}_{\mathsf{K}_{G_{{\ell}}}}}) := 
\begin{cases}
   \bot, & \text{if } P(\mathbf{a}) = False \\
       \mathsf{K}_{G_{\ell}},  & \text{if } P(\mathbf{a}) = True
\end{cases}
\tag{5}\label{eq:5} 
\]
where, the vector $\mathbf{a}$ represents the attributes of the user. 

After the first successful decryption of $c_{P{_\ell}_{\mathsf{K}_{G_{{\ell}}}}}$, the user can proceed with decrypting the actual PSOs by computing
\[\
    \mathsf{PSO}_i := \mathsf{SKE.Dec}(\mathsf{K}_{G_\ell}, c_{\mathsf{PSO_i}})
    \tag{6}\label{eq:6} 
\]
During decryption, only PSOs belonging to the sensitivity groups that the user is authorized to access are decrypted. The corresponding plaintext is then restored into the image, while PSOs from higher-sensitivity groups remain encrypted and inaccessible.

%% file: sections/experimental.tex
\section{Experimental Setup}
\label{sec:Experimental_setup}

This section outlines the experimental setup and the design decisions made to evaluate the proposed architecture. We describe the datasets, implementation details, and evaluation criteria used to assess the performance of each module. All runtime measurements reported in this paper, including ML inference times in ~\ref{subsec:pso_detection} and encryption/decryption times in Table~\ref{tab:runtime}, were performed on a laptop equipped with a 13th Gen Intel(R) Core(TM) i5-1345U
CPU (1.60\,GHz), 32\,GB RAM, and Intel Iris Xe integrated graphics. All experiments were performed using Python 3.10.12, employing Charm-Crypto 0.50 for ABE and cryptography 46.0.1 for symmetric encryption, specifically AES in CBC mode. For training, Kaggle’s T4x2 GPUs were used with PyTorch 2.6. We release the source code upon publication, in which the list of all libraries used can be found.

\subsection{Training Dataset Preparation}



Curating a large-scale dataset of ``private'' images is inherently difficult, since such content is rarely shared publicly. For our use case, images are also required to include various privacy-sensitive objects ({\PSO}s) with different sensitivity scores. Ideally, these scores would be determined through a user study or align with the common perception. Given these challenges, constructing and annotating a new dataset is beyond the scope of this paper. Instead, we examine the suitability of publicly available visual privacy datasets for evaluating our architecture. 

To evaluate the feasibility of our proposed system, we required a dataset that (i) contains multiple PSOs per image to capture contextual relationships, 
(ii) incorporates some notion of scoring that reflects the perceived sensitivity of the PSOs, 
and (iii) includes visually localizable annotations so that PSOs can be spatially pinpointed.
We investigated four large-scale image privacy datasets that contain annotated PSOs: VISPR~\cite{orekondy17iccv}, VISPR-Redactions~\cite{Orekondy2018Connecting}, WizViz~\cite{Gurari2019VizWiz}, BIV-Priv-Seg~\cite{Tseng2025BIV}. Table~\ref{tab:dataset_comparison} provides a comparison of their properties. While VizWiz and BIV-Priv-Seg costly contain a single PSO per image and lack sensitivity scores, VISPR contains private content that is not visually localizable. VISPR-Redactions satisfied all of our requirements and was therefore selected for evaluating our system.

VISPR-Redactions, which is a specially curated subset of VISPR, includes 8,473 images and provides region-level annotations for 24 different {\PSO} classes. Sensitivity scores are inherited from VISPR, where 50 participants rated their comfort level with different object classes on a scale from $1$ (not violated) to $5$ (extremely violated). For each class, the mean score was calculated and linearly normalized to the range $[0.1, 1]$. To improve consistency and class imbalance, we merged some classes (the location class also includes landmark and home address), resulting in a final set of 22 classes grouped into three types: visual, textual and multimodal. 



\begin{table}[t]
    \centering
    \caption{Comparison of visual privacy datasets. (Annotated objects refer to privacy-sensitive objects, and content availability indicates if these objects are visible, or redacted with masking, blurring, replacing with some other content.)}
    \label{tab:dataset_comparison}
        \resizebox{\columnwidth}{!}{%
        \begin{tabular}{l|cccc}
        \toprule
        \multirow{2}{*}{Dataset} & Avg \# of & Content & \multirow{2}{*}{Polygons} & Sensitivity \\
        & Annotated PSOs &  Availability &  &  Score \\
        \midrule
        VISPR~\cite{orekondy17iccv}& 5.2 & \faCircle & \faCircleO & \faCircle\\
        VISPR-Redactions~\cite{Orekondy2018Connecting} & 5.6 & \faCircle & \faCircle & \faCircle\\
        WizViz~\cite{Gurari2019VizWiz} & 1.6 & \faCircleO & \faCircle & \faCircleO \\
        BIV-Priv-Seg~\cite{Tseng2025BIV} & 0.9 & \faCircle & \faCircle & \faCircleO \\
        \bottomrule
    \end{tabular}}
\end{table}

For visual {\PSO}s, segmentation models were trained using the entire training set. Pixel-level annotations ensured that only sensitive regions were detected. For multimodal {\PSO}s, only 1,356 images in the dataset contained at least one relevant object (e.g., credit card, passport, ticket). Including the full dataset would have caused severe class imbalance since most images contained no multimodal {\PSO}. To address this, we retained 1,356 positive samples together with negative samples (images lacking multimodal {\PSO}), producing a focused subset of 4,068 images with an approximate positive-to-negative ratio of 1:2. This subset was randomly split into 60\% training, 20\% validation, and 20\% test sets, and then used to train the object detection models. Also, our CAPC algorithm in the post-correction module required a textual content to facilitate semantic post-corrections. To support this, OCR was applied to the bounding boxes of all annotated multimodal {\PSO}s, resulting in 4,581 text instances paired with their corresponding ground-truth labels. 

For textual {\PSO}s, we constructed a dedicated dataset by applying OCR to training images. The text segments located within the annotated regions were labeled with the corresponding class, while those outside were initially marked as \textit{safe}. To mitigate class imbalance, a portion of the safe samples ($\approx50$k) was randomly removed. This produced 17,233 labeled text instances spanning both sensitive and non-sensitive content. NLP classifiers trained on this dataset feed predictions into the post-correction module before passing structured output to encryption. For evaluation, the same OCR-based procedure was applied to the test split, yielding 6,974 labeled text instances for the test set.



\begin{figure}[t]
    \centering
    \includegraphics[width=1.0\linewidth]{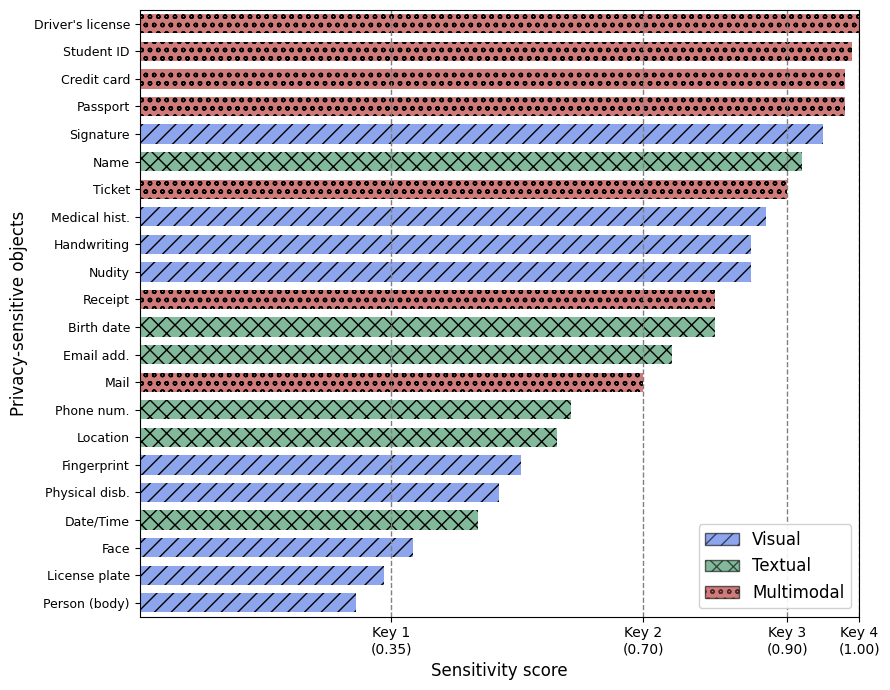}
    \caption{Sensitivity scores of available objects in the dataset and corresponding symmetric keys that can decrypt four different sensitivity groups.}
    \label{fig:privacy_bar3}
\end{figure}

\paragraph{Sensitivity Group Assignment:} As explained in the proposed architecture, System Admin is responsible for defining sensitivity groups and providing this description to other modules in the system plane. Based on object labels and sensitivity scores available from the dataset, we define $L=4$ different sensitivity groups with thresholds: $\{0.35, 0.7, 0.9, 1.0\}$. Figure~\ref{fig:privacy_bar3} illustrates the ranked sensitivity scores of the objects, the sensitivity groups, and the associated symmetric keys that can decrypt these groups.
%
%
The rationale for fixing the number of sensitivity groups to four is two-fold. First, it provides a clear structure for illustrating our solution while being adequately practical to evaluate the performance. Secondly, it aligns with commonly adopted data classification practices, which typically categorize information into four hierarchical levels (e.g., \textit{public}, \textit{internal}, \textit{confidential}, and \textit{restricted}). The number of groups, however, can be adjusted to accommodate different application or system requirements.  

\subsection{Algorithm Benchmarking}

As stated in Section~\ref{subsec:detection_module}, visual {\PSO}s require segmentation to produce precise pixel-level masks to preserve utility. To cover a range of well-known approaches, we fine-tuned four representative segmentation models (with two variants) on the VISPR-Redactions dataset: 1) DeepLabV3+~\cite{chen2018encoder}, an encoder-decoder algorithm that leverages atrous (or diluted) convolutions for extracting features at different resolutions, 2) YOLOv8-Seg (\texttt{s} and \texttt{x} variants)~\cite{yolov8}, an extension of the popular YOLO object detector with mask prediction, 3) SegFormer-B5~\cite{xie2021segformer} which combines hierarchical transformer encoder with a multi layer perceptron decoder, 4) Mask R-CNN~\cite{he2017mask} with ResNet-101-FPN and ResNeXt-101-FPN backbones, which extends Faster R-CNN with a parallel mask prediction branch. We measured the performance of segmentation models by calculating the mean Intersection over Union (mIoU). mIoU is defined as the average ratio of intersection over union between predicted and ground-truth masks across all classes. 



Textual {\PSO}s (e.g., name, email address, birthdate) require semantic interpretation rather than visual cues. Therefore, as stated in Section~\ref{subsec:detection_module}, we first extracted text using OCR and then classified each instance with three transformer-based language models that had fine-tuned over VISPR-Redactions: 1) BERT~\cite{devlin2019bert}, an encoder-only model that learns contextual word representations, 2) DeBERTa~\cite{he2021deberta}, which improves BERT with disentangled attention algorithm and mask decoder, 3) MPNet~\cite{song2020mpnet}, which combines masked and permuted language modeling during training. In addition, we introduced a rule-based correction module (Section~\ref{subsec:correction_module}), Post-BERT, which refines the predictions of BERT based on the contextual cues surrounding the detected text. These rule-based adjustments are listed previously in Section~\ref{subsec:correction_module}. 

Multimodal {\PSO}s (e.g., credit cards, passports, and tickets) in the data set have structured rectangular forms and contain different types of information that must be extracted for accurate identification, thus requiring object detection. We selected four object detection models for their balance in accuracy and speed, and fine-tuned them over VISPR-Redactions: (1) Cascade R-CNN~\cite{cai2018cascade}, which refines predictions through a sequence of progressively selective detecters against false positives, 
(2) RetinaNet~\cite{lin2017focal}, a single-stage detection model designed to handle class imbalance problem with focal loss, 
(3) Faster R-CNN with ResNeXt-101-FPN backbone~\cite{ren2015faster}, two-stage detector that combines region proposals with feature pyramids, 
and (4) YOLOv8~\cite{yolov8}, a widely popular single-stage detection model optimized for real-time performance. 
To further distinguish visually similar objects, we incorporated a Context-Aware Post-Correction (CAPC) module on top of these models: After the detection module, OCR is applied to the bounding box, and the extracted text is used to adjust the predicted label of the identified object, where DeBERTa is used for reclassification. Detection performance is measured by calculating the mean average precision (mAP) when the intersection over union (IoU) threshold varies between $0.5$ and $0.95$, capturing both the localization and the classification accuracy.


%% file: sections/results.tex
\section{Results}
\label{sec:results}

\begin{figure*}[t]
    \centering
    \setlength{\tabcolsep}{2pt} 
    \begin{tabular}{cccccc}

        \includegraphics[width=0.16\textwidth]{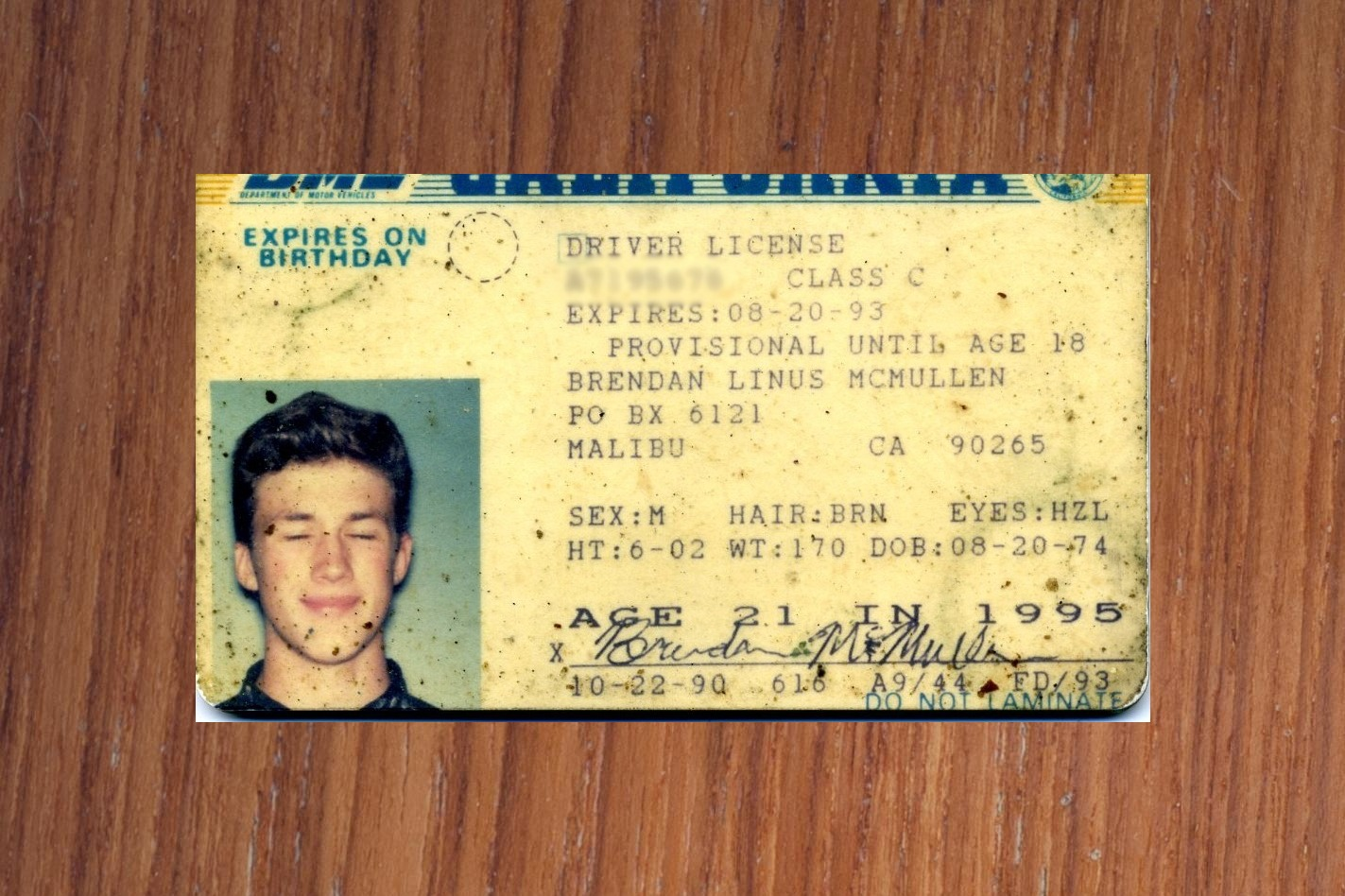} &
        \includegraphics[width=0.16\textwidth]{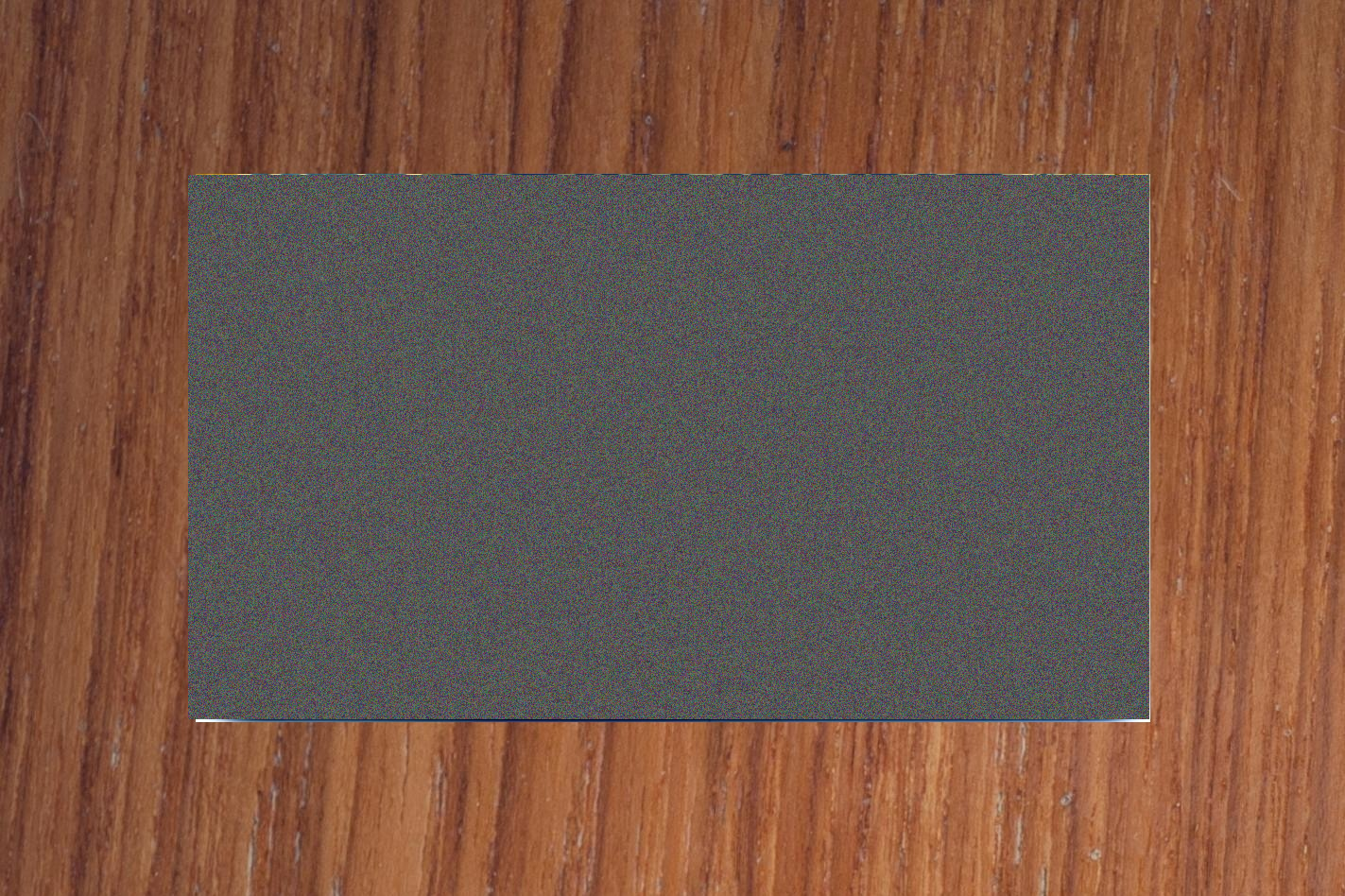} &
        \includegraphics[width=0.16\textwidth]{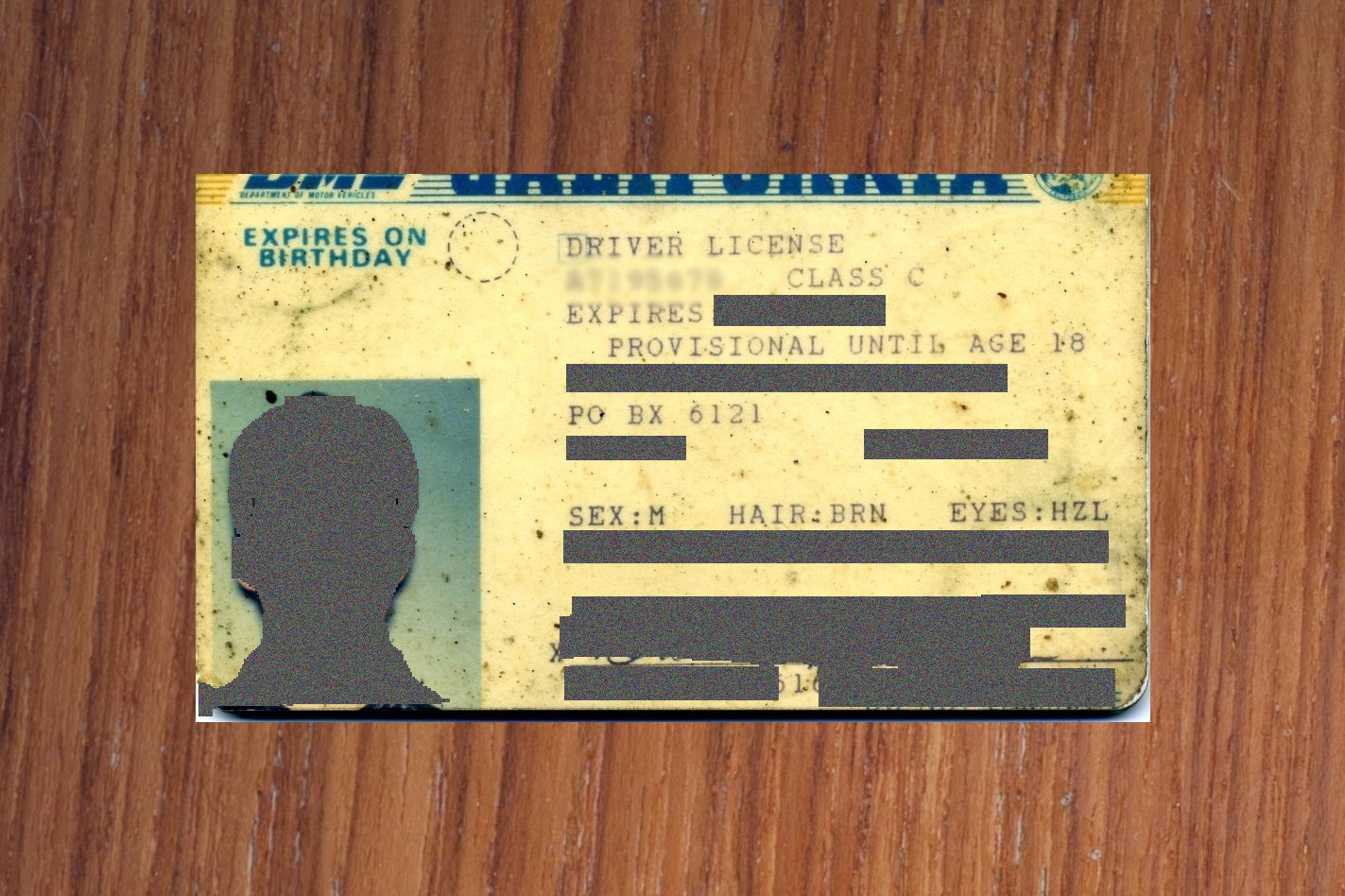} &
        \includegraphics[width=0.16\textwidth]{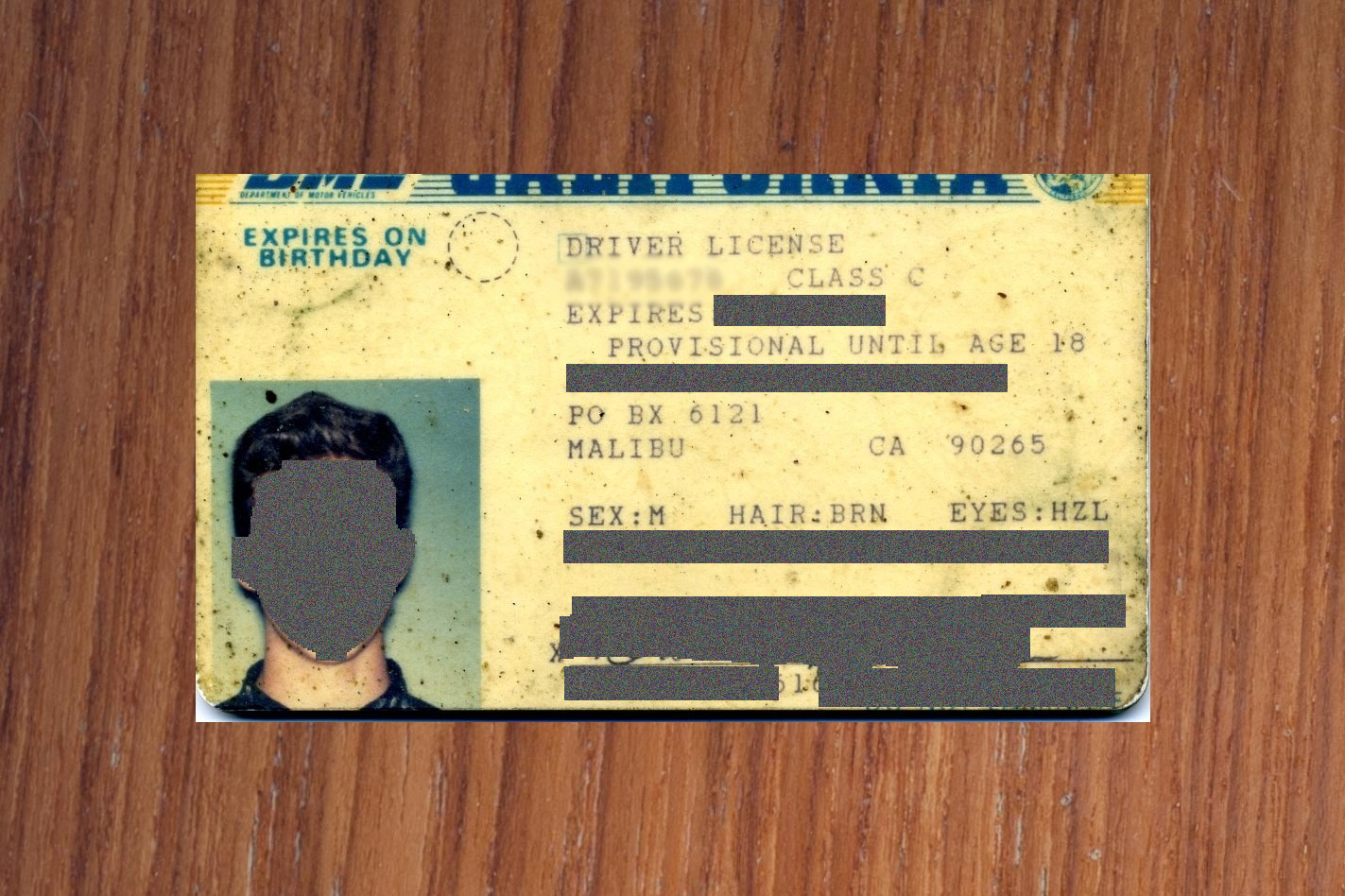} &
        \includegraphics[width=0.16\textwidth]{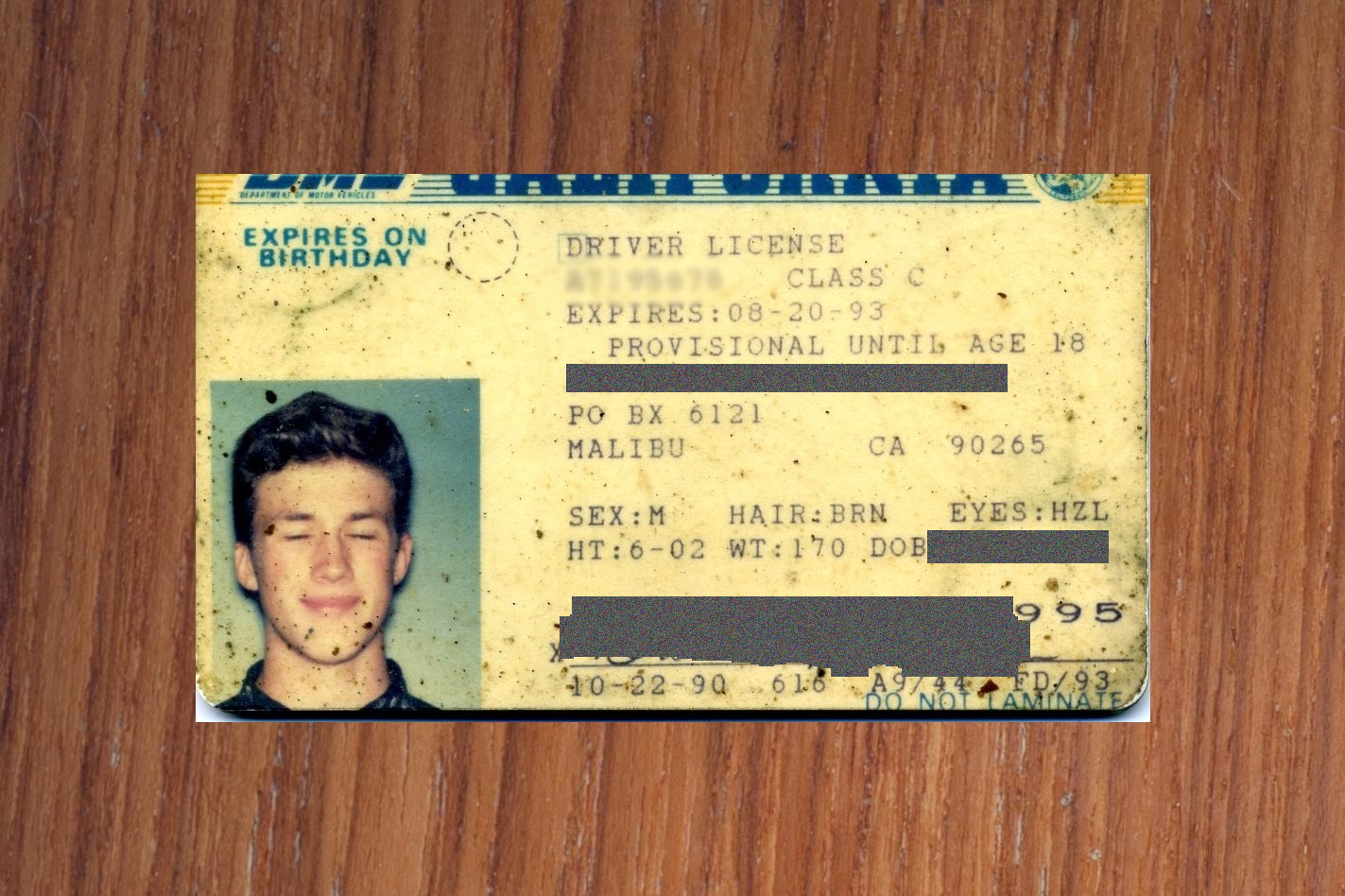} &
        \includegraphics[width=0.16\textwidth]{figures/decrypted_image_abekey4.png}  \\
        \small (a) Original &
        \small (b) Encrypted &
        \small (c) ABE key 1 &
        \small (d) ABE key 2 &
        \small (e) ABE key 3 &
        \small (f) ABE key 4
    \end{tabular}
    \caption{Progressive decryption example under fine-grained access control. Privacy-sensitive objects are grouped into four sensitivity groups with predefined threshold values (see Table~\ref{tab:class_key_scores}) Users holding higher-level ABE decryption keys can gain access to progressively more sensitive content, while unauthorized regions remain scrambled.}
    \label{fig:encryption_results}
\end{figure*}

In this section, we discuss the performance of PSO detection and protection in terms of effectiveness, efficiency, and scalability. 

\subsection{PSO Detection Performance}
\label{subsec:pso_detection}

Table~\ref{tab:all_results_visual} presents the performance of segmentation models, highlighting variations across object categories. SegFormer-B5 and DeepLabV3+ perform particularly well on categories including face, person, handwriting, and fingerprint, but fail in classes that have fewer samples in the training set, such as license plate, physical disability, and medical history. In contrast, YOLO-Seg, FPN-R, and FPN-X provide more balanced performance results in all classes, with FPN-X101 achieving the highest overall mIoU score. Our results indicate that all models exceed a certain threshold for face and person detection, reflecting knowledge already embedded in pre-trained weights and refined during fine-tuning. However, their performance drops significantly for classes such as physical disability, signature, and medicine, largely due to class imbalance. These categories are considered highly sensitive; therefore, fewer samples are publicly available for training. This creates a fundamental challenge: good detection performance is essential, but data scarcity negatively impacts the detection performance. Potential remedies include augmenting datasets with synthetic samples or exploring few-shot approaches (as in ~\cite{Tseng2025BIV})
to enable detection of multiple sensitive regions from limited examples.


Table~\ref{tab:all_results_textual} shows that all three classifiers achieve similar levels of performance, with BERT slightly outperforming DeBERTa and MPNet in overall accuracy and macro-averaged F1 scores. Incorporating our post-correction module with rule-based adjustments further improves the results of BERT. Specifically, Post-BERT refines BERT's predictions using contextual cues, leading to substantial gains in a challenging class birthdate by improving the macro-averaged F1 score from 11\% to 45\%. Although small declines are observed in date, phone, and safe classes, Post-BERT ultimately achieves the best overall and balanced scores, demonstrating that rule-based corrections can effectively complement language models for sensitive text detection.

\begin{table}[h]
\small
\setlength{\tabcolsep}{2.5pt}
\centering
\caption{\normalfont Detection, classification and post-correction results for visual (a), textual (b), multimodal (c) privacy-sensitive objects. \textbf{Bold} and \textit{italicized} results denote the highest and second highest scores in each column, respectively.}

\label{tab:all_results}

\begin{subtable}{\columnwidth}
\footnotesize 
\caption{\normalfont Visual PSOs}\label{tab:all_results_visual}
\begin{tabular}{lcccccccccc}
\toprule
Method & \makecell{mIoU (w)} & \makecell{face} & \makecell{lic\\plt} & \makecell{per\\son} & \makecell{nud\\ity} & \makecell{hnd\\wrt} & \makecell{phy\\dsb} & \makecell{medic\\hist} & \makecell{fing\\prnt} & \makecell{sig\\ntr} \\
\midrule
DLV3+\cite{chen2018encoder}     & 39.3 (66.0) & 65.2 & 0.0 & 70.5 & 22.9 & \textit{58.8} & 0.0 & 0.0 & \textbf{67.2} & 17.7 \\
YOLO-s\cite{yolov8}             & 41.6 (58.7) & 67.8 & 41.3 & 61.4 & 35.4 & 40.1 & \textbf{29.1} & \textit{14.0} & 14.5 & 23.2 \\
YOLO-x\cite{yolov8}             & 43.4 (60.6) & 69.5 & 41.6 & 62.9 & \textit{39.9} & 44.2 & \textit{25.4} & \textbf{18.2} & 19.1 & 22.3 \\
SFM-B5\cite{xie2021segformer}   & 42.1 (\textbf{75.8}) & \textbf{78.6} & 0.0 & \textit{79.4} & \textbf{46.6} & \textbf{66.8} & 0.0 & 0.0 & \textit{55.7} & 0.0 \\
FPN-R\cite{he2017mask}          & \textit{43.8} (71.5) & 73.6 & \textit{44.7} & 78.1 & 32.5 & 29.6 & 17.8 & 12.1 & 30.6 & \textbf{26.6} \\
FPN-X\cite{he2017mask}          & \textbf{46.6} (\textit{74.0}) & \textit{74.1} & \textbf{53.3} & \textbf{80.7} & 34.6 & 32.7 & 24.9 & 11.1 & 37.6 & \textit{26.4} \\
\bottomrule
\end{tabular}
\end{subtable}

\vspace{4pt}

\begin{subtable}{\columnwidth}
\footnotesize 
\caption{\normalfont Textual PSOs}
\label{tab:all_results_textual}
\begin{tabular}{lccccccccc}
\toprule
Method & \makecell{acc.} & \makecell{macro\\avg F1.} & \makecell{name} & \makecell{phone} & \makecell{date\\time} & \makecell{birth\\date} & \makecell{email\\addr} & \makecell{loc} & \makecell{safe} \\
\midrule
MPNet\cite{song2020mpnet}   & 80.4 & 68.1 & 76.5 & 69.2 & 91.4 & 8.6  & 78.3 & 72.3 & 80.4 \\
DeBERTa\cite{he2021deberta} & 81.6 & 70.5 & 76.9 & 74.4 & 91.7 & \textit{13.0} & 80.6 & 76.0 & 80.8 \\
BERT\cite{devlin2019bert}   & \textit{91.1} & \textit{79.8} & \textit{91.3} & \textbf{89.6} & \textbf{94.8} & 11.2 & \textbf{91.7} & \textit{88.7} & \textbf{91.6} \\
\cdashline{1-10}
Post-BERT (Ours)                  & \textbf{91.2} & \textbf{84.2} & \textbf{92.6} & \textit{84.7} & \textit{94.7} & \textbf{45.1} & \textbf{91.7} & \textbf{90.2} & \textit{90.3} \\
\bottomrule
\end{tabular}
\end{subtable}

\vspace{4pt}

\begin{subtable}{\linewidth}
\footnotesize 
\caption{\normalfont Multimodal PSOs}
\label{tab:all_results_multimodal}
\begin{tabular}{lcccccccc}
\toprule
Method & \makecell{mAP} & \makecell{credit\\card} & \makecell{pass\\port} & \makecell{driver\\license} & \makecell{student\\id} & \makecell{mail} & \makecell{receipt} & \makecell{ticket} \\
\midrule
Cascade R-CNN\cite{cai2018cascade}     & 46.2 & 23.5 & 85.2 & 49.8 & 18.1 & 37.4 & 50.9 & 58.7 \\
RetinaNet\cite{lin2017focal}  & 49.9 & \textbf{38.7} & 84.1 & \textit{55.0} & 29.2 & 45.1 & 48.8 & 48.9 \\
Faster R-CNN\cite{ren2015faster}  & 51.1 & 37.5 & 86.1 & 50.8 & 29.6 & 52.9 & 39.8 & 60.9 \\
YOLO\cite{yolov8}             & \textit{58.8} & \textit{38.6} & \textbf{89.7} & 51.6 & \textit{52.4} & \textbf{58.0} & \textbf{51.1} & \textit{70.3} \\
\cdashline{1-9}
CAPC (Ours)                           & \textbf{64.5} & \textit{38.6} & \textbf{89.7} & \textbf{74.5} & \textbf{67.1} & \textbf{58.0} & \textit{46.6} & \textbf{77.9} \\
\bottomrule
\end{tabular}
\end{subtable}

\end{table}

Table~\ref{tab:all_results_multimodal} reports the performance of object detectors on multimodal {\PSO}s. YOLOv8 outperforms Faster R-CNN and Cascade R-CNN, particularly in categories such as ticket and email, which are characterized by their distinctive structures. Therefore, we selected YOLOv8 as the baseline detector for evaluating our Context-Aware Post-Correction (CAPC) method. CAPC further improves YOLO's ability by correctly differentiating between visually similar pairs such as ticket vs. receipt and student ID vs. driver's license, resulting in the highest overall mean Average Precision (mAP) among all evaluated models.

Based on the results presented above, we conjecture that the detection module can benefit from incorporating multiple models to achieve the best performance. Since the type of cues in a given input is unknown in advance, an \emph{integrated ensemble approach} that combines complementary models with post-correction methods is beneficial. As an illustrative example, we construct such an ensemble by selecting the top-performing algorithms benchmarked in Table~\ref{tab:all_results}: Given an input image, OCR is used to extract all textual information. Post-BERT is used to predict labels of the extracted text, integrating rule-based corrections on top of the BERT predictions. Multimodal {\PSO}s are detected with CAPC-enhanced YOLOv8, which refines bounding box predictions with context-aware post-correction. Finally, FPN-X101 extracts visual {\PSO}s, producing pixel-level masks. The average processing time for 2989 test images was measured as 6.01 seconds when executing this flow sequentially and generating the associated metadata.

\subsection{PSO Protection Performance}
\label{subsec:encryption_results}

Figure~\ref{fig:encryption_results} illustrates a case study that demonstrates progressive decryption under our ABE-based access control. In this example, eight privacy-sensitive objects with scores in $[0, 1]$ are placed into four sensitivity groups for better visualization. Table~\ref{tab:class_key_scores} shows the mapping between the sensitivity groups, their corresponding thresholds, and the symmetric keys associated with each group. This example highlights how the proposed architecture enforces fine-grained access to the image, where higher-privileged users progressively unlock more privacy-sensitive content, while unauthorized regions remain \emph{irreversibly} scrambled.



\begin{table}[h]
\centering
\caption{Sensitivity scores of objects from Figure~\ref{fig:encryption_results} and the corresponding symmetric keys. (Sensitivity groups: $0.10 \leq \text{group 1} \leq 0.25$, $0.25 < \text{group 2} \leq 0.50$, $0.50 < \text{group 3} \leq 0.75$, $0.75 < \text{group 4} \leq 1.00$).}
\label{tab:class_key_scores}
\resizebox{0.65\columnwidth}{!}{%
\begin{tabular}{l c c}
\toprule
\textbf{Object} & \textbf{Sensitivity } & \textbf{Symmetric} \\
\textbf{Name} & \textbf{Score} & \textbf{Keys} \\
\midrule
Driver's license & 0.25 &  Key 1 \\
\midrule
Person (body)    & 0.30 &  Key 2 \\
Location         & 0.40 &  Key 2 \\
\midrule
Date/Time        & 0.60 &  Key 3 \\
Face             & 0.70 &  Key 3 \\
\midrule
Birth date       & 0.80 &  Key 4 \\
Name             & 0.85 &  Key 4 \\
Signature        & 0.90 &  Key 4 \\
\bottomrule
\end{tabular}}
\end{table}

Table~\ref{tab:runtime} reports the computational cost for encryption and decryption operations on the image dataset. Encryption accounts only for encrypting images, with separate symmetric keys for each sensitivity group, while decryption time includes both the decryption of the symmetric key and the subsequent decryption of the images. As expected, encryption is more time-consuming because it processes all images and keys. Notably, the difference in decryption time between accessing only the lowest-sensitivity group and all groups is just 0.2 seconds, demonstrating efficient selective access and decryption for the user.

\begin{figure}[t]
    \centering
    \includegraphics[width=0.85\linewidth]{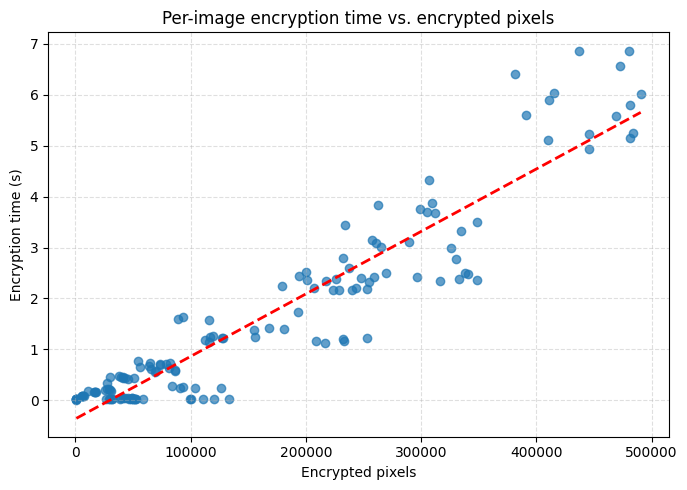}
    \caption{Per-image encryption time vs. total number of encrypted pixels (reported for 350 images randomly selected from the test set).}
    \label{fig:enc_time_vs_pixels}
\end{figure}

Regarding scalability, AES in cipher block chaining mode (AES CBC) operates at a constant time per data block. Consequently, the time required to encrypt image pixels grows linearly with the total number of these pixels, as illustrated in Figure~\ref{fig:enc_time_vs_pixels}. This confirms that encryption scales linearly with the total size of all detected PSOs, ensuring a predictable computational cost when the size of images and possible PSOs are known in advance. 
Finally, Table~\ref{tab:storage_overhead} shows that encryption introduces moderate but consistent storage overhead. On average, the size of an image increases from 4.27 MB to 5.87 MB, adding roughly 1.61 MB per file (38\%). When scaled to 500 images, this corresponds to a total overhead of 0.79 GB. The overhead scales linearly with the number of images and remains within a practical range, making our architecture feasible for real-world deployments, where the overhead can be estimated in advance.

\begin{table}[t]
\centering
\caption{Computational time of encryption and decryption operations on a single image, averaged over the test dataset.}
\label{tab:runtime}
\resizebox{0.82\columnwidth}{!}{%
\begin{tabular}{lc}
\toprule
\textbf{Operation} & \textbf{Avg. Time (s)} \\
\midrule
Encryption   & 11.46 \\
Decryption (For user with ABE key 1) & 0.55 \\
Decryption (For user with ABE key 2) & 0.63 \\
Decryption (For user with ABE key 3) & 0.70 \\
Decryption (For user with ABE key 4) & 0.72 \\
\bottomrule
\end{tabular}}
\end{table}

\begin{table}[t]
\centering
\caption{Storage overhead introduced by encryption.}
\label{tab:storage_overhead}
\resizebox{0.68\columnwidth}{!}{%
\begin{tabular}{lcc}
\toprule
\textbf{Data} & \textbf{Per Image} & \textbf{Total (500 images)} \\
\midrule
Clean      & 4.27 MB   & 2.08 GB \\
Encrypted  & 5.87 MB   & 2.87 GB \\
Overhead    & 1.61 MB   & 0.79 GB \\
\bottomrule
\end{tabular}}
\end{table}



%% file: sections/security.tex
\section{Security Discussion}
\label{sec:security}

We discuss how the proposed design addresses confidentiality and enforces access control while considering practical deployment. 

\paragraph{\textbf{Preserving Confidentiality.}} 
The CryptoCore module encrypts each PSO with a symmetric key tied to its sensitivity group. Each symmetric key is further protected via ABE under policies defined by the AccessPolicy Module. Raw PSOs never leave the CryptoCore module; only metadata is visible to other system components.

\paragraph{\textbf{Access Control.}} 
User access is determined by ABE keys encoding their attributes. Decrypting a symmetric key allows access only to PSOs permitted by the corresponding policy, while higher-sensitivity PSOs remain encrypted, ensuring strict enforcement of access rules.

\paragraph{\textbf{Collusion Considerations.}} 
Low-risk collusions, like users sharing decrypted PSOs, do not expose sensitive content. High-risk collusions involving the AccessPolicy or CryptoCore modules could compromise confidentiality but are out of scope of this work. However, Trusted Execution Environments could potentially be used to mitigate them.

\paragraph{\textbf{Metadata Leakage}}  
Although raw PSOs are encrypted, metadata may leak side-channel information. We define a \emph{leakage function} $\mathcal{L}$ mapping a dataset $\mathcal{D}$ to potentially exposed information:
\[
\mathcal{L}(\mathcal{D}) =
\left\{
\begin{aligned}
&\text{PSO positions,class frequencies},\\
&\text{confidence distributions,sensitivity group counts},\\
\end{aligned}
\right\}.
\]

This leakage is inherent to object-detection-based systems and unavoidable without obfuscation~\cite{barak2001possibility} or oblivious RAM (ORAM)~\cite{goldreich1996software}, which are impractical in this setting. In practice, ORAM adds significant bandwidth and latency overhead due to repeated oblivious memory accesses, and efficient general-purpose obfuscation does not exist beyond theoretical constructions, making both incompatible with real-time object detection pipelines. 

To formalize the confidentiality guarantees and access control guarantees, we show that even with access to $\mathcal{L}(\mathcal{D})$, no adversary can gain non-negligible advantage in recovering PSO contents without satisfying the associated access policies.

\begin{theorem}[Confidentiality \& Access Control under Metadata Leakage]
If $\mathsf{SKE}$ is IND-CPA~\footnote{IND-CPA: Indistinguishability under Chosen-Plaintext Attack, a standard notion of semantic security.} secure and $\mathsf{ABE}$ is IND-CPA secure and collusion-resistant, then no PPT adversary $\mathcal{A}$ that cannot satisfy policy $P_\ell$ and observes metadata via $\mathcal{L}$ can distinguish encryptions of two chosen $\mathsf{PSO}$s from group $\ell$ with non-negligible advantage.
\end{theorem}

\begin{proof}[Proof Sketch]
Suppose a PPT adversary $\mathcal{A}$ has a non-negligible advantage $\epsilon$ in distinguishing $\mathsf{PSO}_0$ and $\mathsf{PSO}_1$, possibly using $\mathcal{L}(\mathcal{D})$. We construct a PPT adversary $\mathcal{B}$ that breaks $\mathsf{SKE}$ or $\mathsf{ABE}$:

\begin{itemize}
    \item If $\mathcal{A}$ never recovers the symmetric key $\mathsf{K}_\ell$, distinguishing $c_{\mathsf{PSO}_b} \leftarrow \mathsf{SKE.Enc}(\mathsf{K}_\ell, \mathsf{PSO}_b)$ gives $\mathcal{B}$ an IND-CPA attack on $\mathsf{SKE}$ with advantage $\epsilon$.
    \item If $\mathcal{A}$ obtains $\mathsf{K}_\ell$ without satisfying $P_\ell$, $\mathcal{B}$ uses this to distinguish $c_{P_\ell}$ under $\mathsf{ABE}$ IND-CPA, breaking $\mathsf{ABE}$ with advantage $\epsilon$.
    \item Metadata from $\mathcal{L}(\mathcal{D})$ reveal, at most positions, class frequencies, or sensitivity counts. Since it does not expose plaintext PSOs or symmetric keys, it contributes only negligible advantage.
\end{itemize}

Thus, in all cases, $\mathcal{B}$ contradicts the assumed security of $\mathsf{SKE}$ or $\mathsf{ABE}$, and metadata leakage is negligible. Therefore, $\epsilon$ is negligible.

\end{proof}

\paragraph{\textbf{PSO Leakage due to Detection Misses}}
As shown in Section~\ref{sec:results}, ML models can still produce false negatives, i.e., fail to detect certain PSOs, even when an integrated ensemble approach is used. Achieving zero false negatives is challenging with current general-purpose segmentation and object detection models. Therefore, sensitive objects or pixels may remain unencrypted, potentially leading to privacy leakage. To mitigate this limitation, we propose to enable user control over the detection and classification module. In the context of the proposed architecture, this functionality can be incorporated into the Post-Correction Module. Such user interaction and correction mechanisms can support the refinement and further fine-tuning of the ML models, thereby improving control over privacy leakage. The integration of human feedback into the architecture, as well as the development of human validation methods to assess privacy risks due to false negatives, is left for future work. We consider this direction crucial for enhancing the practical applicability of the proposed architecture and for increasing user trust in automated access control systems.

%% file: sections/conclusion.tex
\section{Conclusion}

In this paper, we present a system architecture for fine-grained, policy-driven access control over visual datasets containing privacy-sensitive objects (PSOs). Our solution combines automated PSO detection, post-correction, and a hybrid cryptographic protection scheme to enable selective encryption and secure sharing of sensitive content. The experimental results demonstrate the efficiency and scalability of our solution. Overall, our work provides a practical approach for combining ML-based sensitive-region detection with cryptographic protection and enforcement of access control.

Our work addresses some of the long-standing limitations of traditional access control systems. First, traditional access control systems rely heavily on static policy assignments and manual data classification, both of which are error-prone due to inconsistent and subjective human judgments. Our system overcomes this limitation by introducing semantic adaptability, where ML-based detection helps learn what to protect and how to classify content by interpreting visual, textual, and spatial cues in context. Such data-driven adaptability offers dynamic automation capabilities that, alongside eliminating human errors and inconsistencies, can evolve  with content and context. Second, traditional access control is enforced superficially on the data (i.e., around the data rather than within it) through file-system permissions, identity services, and application logic. Such an approach fails when data leaves its origin, e.g., when users share files outside a controlled environment or manually copy its content. Once detached from its enforcement layer, the sensitive information becomes exposed and unguarded. 
To address this problem, our design binds access rules to the data itself, ensuring that protection travels with the data and remains effective in situations where an access-protected data accidentally shared with an unauthorized user outside the administrative domain or the system's trusted boundaries. 
By combining semantic adaptability and cryptographic binding, our work points towards a new design paradigm for secure system architecture. 

While this paper focuses on image datasets, the concepts and architecture we propose are modality-agnostic. Thus, our work can be naturally extended to more diverse data forms, such as audio, video, or sensor feeds. Moreover, the modular design allows personalization of access control policies (e.g., based on user roles, devices, or system environment) that can adapt dynamically to real-world usage conditions. The generalizability and modularity of our design further enhance its applicability and prospects beyond visual data, positioning this work as a step toward building 
a smart access control solution for modern data ecosystems.